\documentclass[a4paper]{article}

\usepackage{INTERSPEECH2020}
\usepackage{xcolor}
\usepackage[show]{inotes}
\usepackage{multicol}
\usepackage{multirow}
\usepackage{graphicx}
\usepackage{hyperref}
\usepackage{enumitem}
\usepackage{arydshln}
\setlist[itemize]{leftmargin=*}
\usepackage{varwidth}
\usepackage{bigstrut}
\usepackage{paralist}

\title{Word Error Rate Estimation Without ASR Output: e-WER2}
\name{Ahmed Ali$^{1}$, Steve Renals$^2$}
\address{
  $^1$Qatar Computing Research Institute, HBKU, Doha, Qatar\\
  $^2$Centre for Speech Technology Research, University of Edinburgh, UK}
\email{amali@hbku.edu.qa, s.renals@ed.ac.uk}

\begin{document}

\maketitle
\begin{abstract}
Measuring the performance of automatic speech recognition (ASR) systems requires manually transcribed data in order to compute the word error rate (WER), which is often time-consuming and expensive. In this paper, we continue our effort in estimating WER using acoustic, lexical and phonotactic features. Our novel approach to estimate the WER uses a multistream end-to-end architecture. We report results for systems using internal speech decoder features (glass-box), systems without speech decoder features (black-box), and for systems without having access to the ASR system (no-box). The no-box system learns joint acoustic-lexical representation from phoneme recognition results along with MFCC acoustic features to estimate WER. Considering WER per sentence, our no-box system achieves 0.56 Pearson correlation with the reference evaluation and 0.24 root mean square error  (RMSE) across 1,400 sentences. The estimated overall WER by e-WER2 is 30.9\% for a three hours test set, while the  WER computed using the reference transcriptions was 28.5\%.
\end{abstract}

\section{Introduction}
Automatic Speech Recognition (ASR) has accomplished great success, primarily due to advances in the end-to-end neural networks and the modular hybrid HMM-DNN architectures. As a result, the quality of ASR has improved dramatically, leading to growing adoption in personal assistant devices, smart phones and broadcast media monitoring. Despite this progress, ASR performance is still closely tied to how well the training data matches the test conditions, such as the variability of different microphones or background noises. While researchers in \cite{xiong2016achieving,saon2017english} discussed achieving human parity on conversational speech, recent competitions in ASR  \cite{watanabe2020chime,aliMGB3} reported considerably poorer results due to dialectal speech, 
simultaneous recordings from multiple microphone arrays, and  background noise. 

Word Error Rate (WER) is the standard approach to evaluate the performance of a large vocabulary continuous speech recognition (LVCSR) system. To obtain a reliable estimate of the WER, at least two hours of manually transcribed test data is typically required -- a time-consuming and expensive process. It is, thus, of interest to develop techniques which can automatically estimate the quality of the ASR.

Such quality estimation techniques have been extensively investigated for machine translation \cite{fan2019bilingual, yang2019ccmt, fonseca2019findings}, with extensions to spoken language translation \cite{ng2015study,ng2016groupwise}.  Although there is a long history of exploring word-level confidence measures for speech recognition \cite{evermann2000posterior,cox2002high,jiang2005confidence,seigel2011combining,huang2013predicting, kalgaonkar2015estimating, del2018speaker, swarup2019improving}, there has been fewer attempts on the direct estimation of speech recognition errors \cite{seigel2014detecting,simonnet2017asr}.
\newline
\noindent Previously, we proposed e-WER~\cite{ali-ewer}, a method to estimate the total number of errors per utterance ($\hat{ERR}$)  and the total number of words in the reference ($\hat{N}$)  as shown in section \ref{subsubsec:ewer}. However, that work  assumed having access to a graphemic speech recognition for the predicted language and being able to see the ASR transcription . In this paper, we extend this work by deploying an end-to-end multistream architecture to predict the WER per sentence using language-independent phonotactic features.
Our novel system is able to learn acoustic-lexical embeddings to estimate the error rate directly without having access to the ASR results nor the ASR system -- this is our ``no-box'' WER estimation method, e-WER2\footnote{\url{https://github.com/qcri/e-wer}}.



\section{Related Work}
Several studies have explored estimating the WER in LVCSR. TranscRater  \cite{negri2014quality, de2015multitask, jalalvand2015stacked, jalalvand2015boosted,jalalvand2015driving} estimated the WER per utterance using a large set of extracted features (not including ASR decoder features) to train a regression model (e.g., extremely randomised trees). This work did not report WER estimates for complete recordings or test sets, although it is possible that this could be done using utterance length estimates.


Fan et al \cite{fan2019neural} proposed a novel neural zero-inflated model to predict the WER of the ASR result without transcripts.  They deployed a bidirectional transformer language model conditional on speech features (speechBERT). They adopted the pre-training strategy of token level mask language modeling for speech-BERT as well, and further fine-tune with zero-inflated layer for the mixture of discrete and continuous outputs. They reported results in WER prediction using the metrics of Pearson correlation and mean absolute error (MAE).

Vyas et al \cite{ vyas2019analyzing} used dropout in a novel framework to model uncertainty in prediction hypotheses. They systematically exploited this uncertainty in the output of the acoustic models through the Monte Carlo sampling of the neural networks using dropout at the test time. They were able to estimate the WER without the need for explicit transcription. However, the models must have access to the ASR models to model the uncertainty in the prediction.

\subsection {e-WER} \label{subsubsec:ewer}
In this section, we give a brief overview of our previous e-WER framework \cite{ali-ewer}. We used two speech recognition systems; a word-based LVCSR system and a grapheme-sequence based system. Following \cite{tam2014asr}, we assumed that when two corresponding ASR systems disagree on a sentence or part of a sentence, there is a pattern of the error to be learned. The e-WER architecture also benefits from utterance-based LVCSR internal decoder features
. The e-WER approach is looking for the overall error pattern and not particularly concerned with the error. We directly estimated the numerator in the WER, which is the summation of insertion, deletion and substitution errors, which we refer to as $\hat{ERR}$, the estimated total number of errors per utterance.  We also directly estimated $\hat{N}$, an estimate of the total number of words in the reference as shown in \ref{eq:ewer}. The e-WER predicts two values for each utterance: $\hat{ERR}$ and $\hat{N}$. More details about the e-WER can be found here \cite{ali-ewer}. We use the e-WER system as a baseline reference for this paper.

\begin{equation}
  \mbox{e-WER}=\frac{\hat{ERR}}{\hat{N}}\times 100 \%
\label{eq:ewer}
\end{equation}

\section{e-WER2 Framework} \label{sec:eWER}
In this paper, we develop models to predict the WER per sentence rather than $\hat{ERR}$ or $\hat{N}$. WER per sentence can be scaled by the corresponding sentence duration to calculate the overall e-WER2.

\subsection{Features} \label{subsec:features}
We combine features from the word-based LVCSR system with features from the  phoneme-based system. We split the studied features into the following four groups: 

\begin{compactitem}
\item \emph{L}: lexical features -- the word sequence extracted from the LVCSR;
\item \emph{P}: phoneme features -- the phonotactic sequence extracted from the phoneme recognition, see \ref{subsubsec:phoneme};
\item \emph{D}: decoder features -- total frame count, average log-likelihood, total acoustic model likelihood, and total language model likelihood; and
\item \emph{A}: acoustics features -- the MFCC features are extracted by segmenting each utterance into 25 ms long frames with a 10 ms shift. A Hamming window is applied and the FFT with 512 points is computed. Then, we compute the logarithmic power of 26 Mel-frequency filter-banks over a range from 0 to 8 kHz. Finally, a discrete cosine transform (DCT) is applied to extract the first 13 MFCCs.
\end{compactitem}

\subsubsection{Phonotatic features} \label{subsubsec:phoneme}
In our phonotactic systems, we use Arabic and non-Arabic phone recognizers. For the Arabic recogniser, the HMM-LSTM based acoustic model is trained using 1,200 hours of training data from the MGB-2 datasets \cite{najafian2018exploiting} . In addition to the Arabic
recognizer, we used a phone recognizer from a toolkit developed by Brno University of Technology \cite{matejka2005phonotactic}, trained on Hungarian, but empirically observed to be applicable to multiple languages.  This phone recogniser is based on a long temporal context, and has been widely used to discriminate between various languages and dialects. The intuition for using this system is that a robust phone recogniser is capable of extracting an
accurate phonotactic pattern for the recognised language. We benchmarked Hungarian results against Arabic and the results were similar, thus we decided to deploy this phone recognition system for the multilingual extraction of phonotactic features.

\subsection{Modelling} \label{subsec:modelling}
In this study, we consider four different streams; numerical (\emph{D}), lexical (\emph{L}), phonotactic (\emph{P}) and finally acoustic features (\emph{A}).
\subsubsection{Numerical modelling: (\emph{D})}  \label{subsubsec:numerical-modelling}
We deploy a feed-forward neural network for the numerical features with fully-connected hidden layers (ReLU activation function), with 64 neurons in the first layer and 32 neurons in the second layer followed by a softmax layer with mean squared error loss function. We use dropout rate between layers 0.2, minibatch size of 32 and the number of epochs was up to 50 with an early stopping criterion.
\subsubsection{Acoustic modelling: (\emph{A})}  \label{subsubsec:acoustic-modelling}
We employed deep CNN models, each with five layers where four of them are CNN layers. The same dropout rate, batch size and number of epochs was used as above. Table  \ref{tab:CNN_MFCC_MODEL} shows details of the deployed models. More details about this model can be found here \cite{8683632}.
\begin{table}[htbp]
  \centering
  \caption{The acoustic features deep CNN architecture.}
    \begin{tabular}{r|l|p{14.5em}}
    \multicolumn{1}{l|}{Layer} & Type  & \multicolumn{1}{l}{Details} \bigstrut\\
    \hline
    1     & Conv  & 500 filters + Relu + Stride=1 + kernal wdith=5 \bigstrut\\
    \hline
    2     & Conv  & 500 filters + Relu + Stride=2 + kernal wdith=7 \bigstrut\\
    \hline
    3     & Conv  & 500 filters + Relu + Stride=2 + kernal wdith=1 \bigstrut\\
    \hline
    4     & Conv  & 500 filters + Relu + Stride=1 + kernal wdith=1 \bigstrut\\
    \hline
    5     & MaxPool1D & Global \bigstrut\\
    \hline

    \end{tabular}%
  \label{tab:CNN_MFCC_MODEL}%
\end{table}%

\subsubsection{Textual modelling: (\emph{L} and \emph{P})} \label{subsubsec:textual-modelling}
We use CNN models for phoneme and text processing. The input word sequences are trimmed to a maximum of 100 words for the long sentences, and we padded shorter sentences with zeros. This was followed by an embedding layer of a dimension of 256. Followed by three convolutional layers in parallel to each other with the same number of filters: 512 each, and ReLU activation function. The filters’ sizes were different for each convolution layer: 3, 4 and 5, respectively. The three-convolutional layers were then merged into a single tensor. The same CNN is used for the phoneme sequence. However, a maximum of 200 dimensions were used for phoneme-based CNN. More details about this model can be found 
here \cite{ ali2018multi}.

\subsubsection{Multistream system: \label{subsubsec:no-box}}
\begin{figure*}[t]
\centering
\includegraphics [scale=0.5]{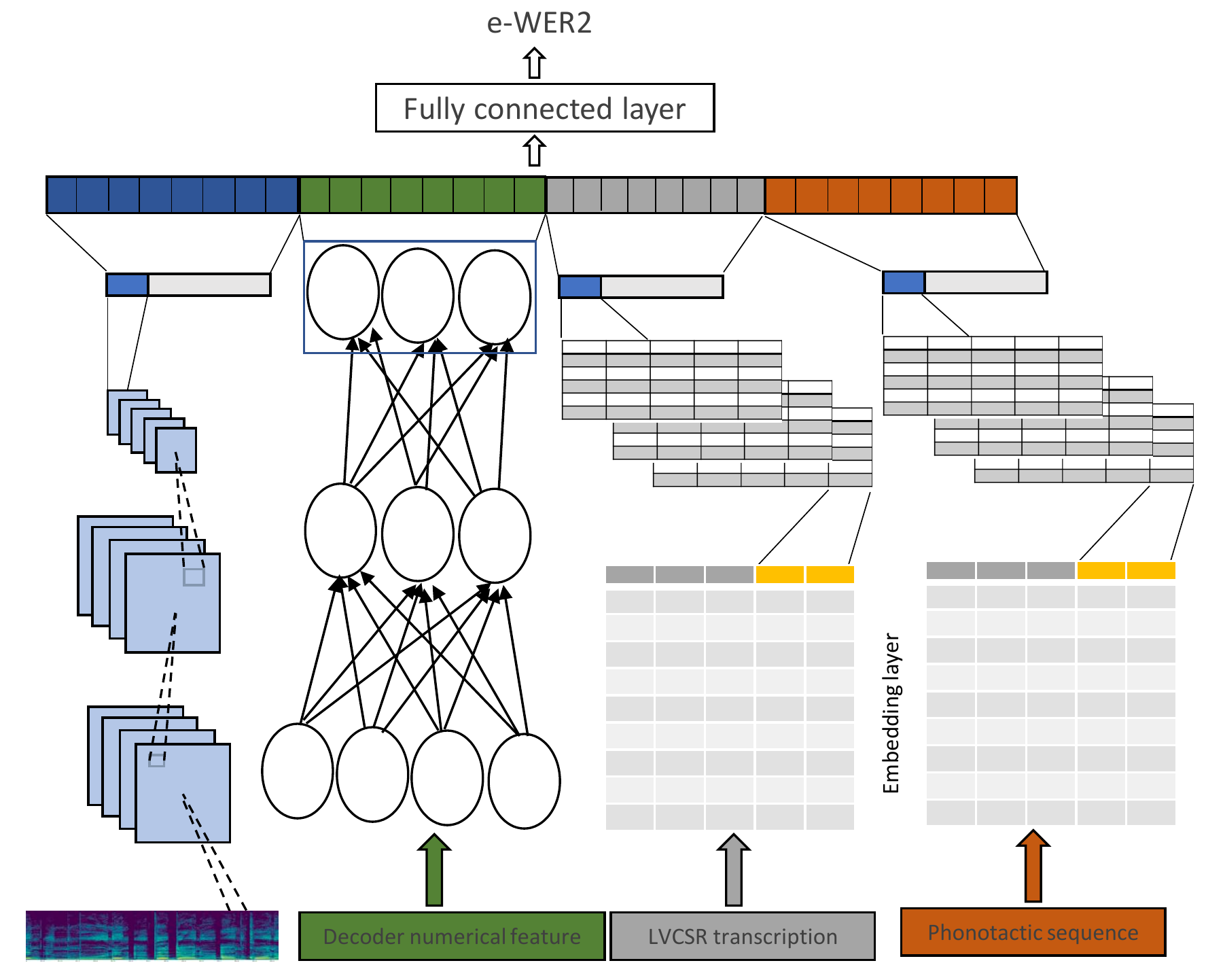}
\caption{
Multistream model architecture of e-WER2. Based on a combination of four features: decoder, acoustics, textual and phonotactics features.}
\vspace{-0.3cm}
\label{MM_system}
\end{figure*}

We combine the four streams: lexical, phonotactic, acoustics and numerical features into a single end-to-end network to estimate word error rate directly. We jointly train the multistream network and their final hidden layers are concatenated to obtain a joint feature space in which another fully connected layer with 32 neurons and Relu activation function to estimate the WER directly. Figure \ref{MM_system} shows the architecture of the multistream approach developed in this paper.

\section{Speech Recognition System}
\label{sec:asr_train}
The LVCSR system is trained using the second Multi-Genre Broadcast challenge data, MGB-2 \cite{ali2016mgb}. The data comprised recorded programs over 10 years of the Aljazeera Arabic TV channel with a total of 1,200 hours of audio that could be used for the acoustic model (AM). The original transcription has no timing information and is not verbatim, having been generated as closed captions for viewers; the quality of the transcription varies significantly. We, therefore, use lightly supervised alignment algorithms in order to recover the timing information for each word. 

For language modelling, we use 130M words crawled from the Aljazeera Arabic website from the period 2000--2011 (background text), as provided for the MGB-2 challenge. 
We have used the provided Buckwalter\footnote{Buckwalter is a one-to-one mapping allowing non-Arabic speakers to understand Arabic scripts, and it is also left-to-right, making it easy to render on most devices.} format for the transcription as well as for the background text.
 LM experiments used a grapheme lexicon of 1.3M words. The grapheme based lexicon has a 1:1 word-to-grapheme mapping, which means the vocabulary size is the same as the lexicon size. 
More details about the data can be found in tables \ref{tab:am_data} and \ref{tab:lm_data}.

\begin{table}[t]
\centering 
\caption{\textit{Data used for acoustic model training, development and evaluation}}
\label{tab:am_data}
\begin{tabular}{@{}llll@{}} \toprule Type & Hours & Programs & \#segments\\
\midrule
Training  & 1,200h & 2,214 & 370K \\
Development & 10h & 17 & 5,800 \\
Evaluation &  10h & 17 & 5,600 \\
\bottomrule \end{tabular}
\end{table}

\begin{table}[t]
\centering
\caption{\textit{In-domain data refers to the training transcripts and Background data refers to the extra Arabic language modeling text provided for the challenge}}
\label{tab:lm_data}
\begin{tabular}{@{}lll@{}} \toprule Type & Tokens & Vocab\\
\midrule  In-domain & 8M &  200k \\
Background & 130M &  1M \\
 \bottomrule \end{tabular}
 \end{table}

\noindent \textbf{Acoustic modelling}: There are many architectures used for the hybrid HMM neural acoustic modeling, with a recent trend in ASR modeling combining different types of layers. Peddinti et al \cite{peddinti2015time} explored using dropout to improve generalisation in DNN training. They reported that combining a time delay neural network (TDNN) with long short term memory  (LSTM) layers outperformed bidirectional LSTM (BLSTM) acoustic modelling. We adopt this architecture. The TDNN-LSTM model consists of 5 hidden layers, each layer containing 1,024 hidden units. We use purely sequence trained neural networks using lattice-free maximum mutual information (LF-MMI) \cite{povey2016purely}. Acoustic models are built using Kaldi ASR toolkit \cite{povey2011kaldi}.

\noindent \textbf{Language modelling}: We train two \textit{n}-gram LMs: a big four-gram LM (bLM$4$), trained using the spoken transcripts and the background text as shown in table \ref{tab:lm_data}; and a smaller four-gram LM (sLM$4$) obtained by pruning bLM$4$ using pocolm\footnote{https://github.com/danpovey/pocolm}. The small LM is used for first-pass acoustic decoding to generate lattices. These lattices
are then rescored using the bLM$4$. 

\section{Data} \label{sec:data}

In our study, we use the same data as \cite{ali-ewer} to benchmark our results. The e-WER2 training and development data sets are the same as the Arabic MGB-2 development and evaluation sets \cite{ali2016mgb}, which is comprised of audio extracted from Al-Jazeera Arabic TV programs recorded by Brightcove in the last months of 2015. They each comprise 10 hours of audio that were not used in the MGB-2 training data. (Other episodes of the same program may have been included in the training set).  To test whether our approach generalises to test sets from a different source,  and not tuned to the MGB-2 data set, we validate our results on another three hours test set collected by  BBC Media Monitoring from different broadcasters during November 2016, as part of the SUMMA project
\footnote{http://summa-project.eu}.  The SUMMA data is  referred to as the test set. All data were manually segmented and labeled. Table \ref{ewer_data} shows more details about the data used for these experiments.

\begin{table}[t]
\center
\caption{Analysis of the train, dev and test data.}
\label{ewer_data}
\hspace*{-0.5cm}
\scalebox{1.0}{
\begin{tabular}{l|l|l|l}
 &  Train & Dev & Test \\
\hline
\# of programs in corpus & 17 & 17 & 24 \\
Utterances  & 5.6K & 5.8K & 1.4K \\
Duration (in hours) & 10.2 & 9.9 & 3.2 \\
2-20 words sentences & 95\%& 96\%& 96\%\\
\textbf{Word count ($N$)} & \textbf{69K} & \textbf{75K} & \textbf{20K} \\
ASR word count (hyp) & 60K & 58K & 18K\\
WER & \textbf{33.1\%} & \textbf{42.6\%} & \textbf{28.5\%}\\
Sentence Error Rate (SER) & 89.1\% & 88.7\% & 86.0\%\\
Total INS & 1.8K & 1.9K & 130\\
Total DEL & 10.2K & 19.1K &2.6K \\
Total SUB & 10.8K &11.1K &2.9K\\
\textbf{ERR count ($ERR$)} & \textbf{22.8K} &\textbf{32.1K} & \textbf{5.7K}\\
\hline
\end{tabular}
}
\end{table}

\section{Experiments and discussions} \label{sec:exp}
We train our end-to-end system to estimate WER per sentence as regression problem. The hyper-parameters for the system were tuned using two evaluation metrices: Pearson correlation and root mean square error (RMSE) for the development set and results are reported for the test set. In our feature ablation study, we evaluated the six following systems:

\begin{compactitem}
\item $\mathcal{A}$: Decoder features + MFCC + lexical features 
\item $\mathcal{B}$: $\mathcal{A}$ + phonotactic features 
\item $\mathcal{C}$: MFCC + lexical features
\item $\mathcal{D}$: $\mathcal{C}$ + phonotactic features
\item $\mathcal{E}$: MFCC
\item $\mathcal{F}$: $\mathcal{E}$ + phonotactic features
\end{compactitem}

\begin{table}[t]
\centering
\caption{Pearson correleation and RMSE report per system. The overall WER reported in \%, the reference overall WER is 28.5\%}
\label{Pearson_RMSE}
\hspace*{-5mm}
\scalebox{1.0}{
\begin{tabular}{c|cc|c}
& \multicolumn{2}{c|}{\bf WER Per Sentence}&{\bf Overall WER} \\
\hline
&Pearson&$RMSE$&e-WER\\
Glass-box baseline &0.8&0.17&26.5 \\
Black-box baseline &0.66&0.35&28.6 \\
\hdashline              
$\mathcal{A}$&0.79&0.2&28.0 \\
$\mathcal{B}$ \footnotesize	{e-WER2 glass-box} &\textbf{0.81}&\textbf{0.18}&\textbf{27.7} \\
$\mathcal{C}$&0.68&0.22&35.3 \\
$\mathcal{D}$ \footnotesize{e-WER2 black-box} &\textbf{0.72}&\textbf{0.2}&\textbf{22.4} \\
$\mathcal{E}$&0.11&0.28&46.1\\
$\mathcal{F}$ \footnotesize	{e-WER2 no-box} &\textbf{0.56}&\textbf{0.24}&\textbf{30.9}\\
\hline
\end{tabular}
}
\end{table}

The first two rows in table \ref{Pearson_RMSE} show the glass-box and black-box results from our baseline system; the e-WER where we combined word-based and grapheme-based ASR results for the same sentence. system $\mathcal{A}$ shows our first multistream architecture which combines acoustics, lexical and decoder features. System ($\mathcal{B}$ \textbf{e-WER2 glass-box}) achieves Pearson correlation of 0.81, which outperforms the glass-box in e-WER with no need to run grapheme based speech recognition for the same language. System $\mathcal{C}$ is trained using the lexical and acoustics features only. System ($\mathcal{D}$ \textbf{e-WER2 black-box}) achieves Pearson correlation of 0.68, which outperforms the e-WER black-box reference systems. 
At this stage, we are confident that our multistream system is capable of learning a joint representation for acoustics and linguistic (textual and phonotactic) features to estimate the WER. 

Our experiments show that the proposed multistream architecture can estimate the WER efficiently without requiring graphemic recognition. Fan et al~\cite{fan2019neural} estimated  WER without the requiring explicit transcriptions, but did require access to the acoustic models.  Here, we ask; Can we estimate the WER without having access to neither the transcript nor the speech recognition system? In our attempt to answer this, we build system $\mathcal{E}$ in table \ref{Pearson_RMSE} which uses only acoustic features. Clearly, the system is not capable of learning any pattern given that the MFCC features. When combined with phoneme recognition output ($\mathcal{F}$) we see a large improvement in Pearson correlation by combining acoustic and phonotactic features, still without access to the ASR system. 

\begin{figure}[t]
\centering
\includegraphics [scale=0.6]{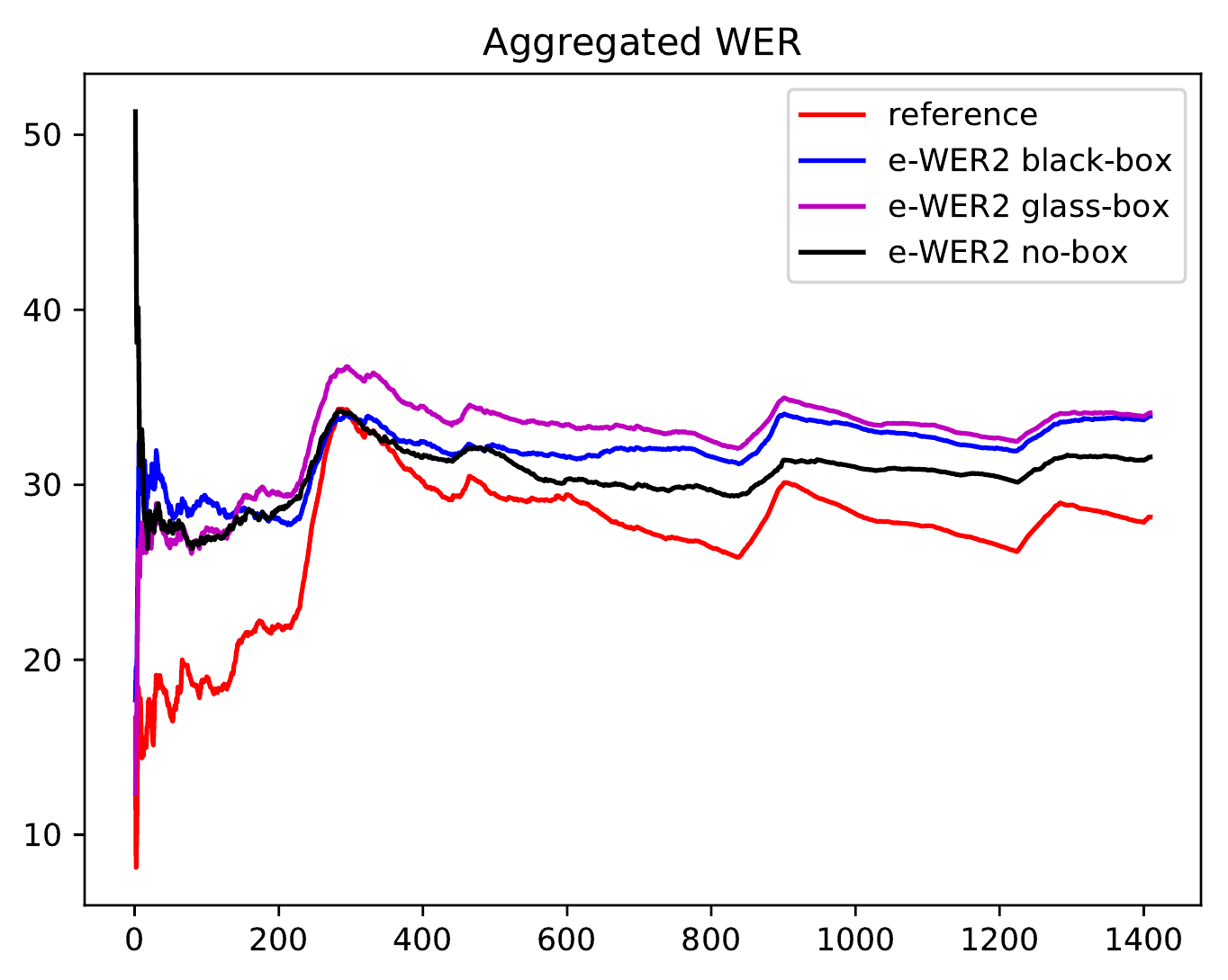}
\caption{Test set cumulative WER over all sentences \\
X-axis is duration in hours and Y-axis is WER in \%.}
\vspace{-0.3cm}
\label{wer_acc}
\end{figure}

To further visualise these results, figure~\ref{wer_acc} plots the cumulative WER, glass-box, black-box and no-box in the e-WER2 framework, across the three hours test set. The large difference during the first 100 utterances arises owing to the glass-box and black-box systems in the e-WER2 framework are capable of better estimation with fewer data points. 
It is worth to mention that when we swap train and dev, the results are similar. 

\section{Conclusions} \label{sec:conclusion}
This paper continues our effort in predicting speech recognition WER without requiring a gold-standard reference transcription. We presented an end-to-end multistream based regression model to predict the WER per sentence. Our approach benefits from combining word-based and phoneme-based ASR results, in addition to the MFCCs for the same sentence. Our experiments indicate that this approach can effectively estimate WER per sentence and we have aggregated the estimated results to predict WER for complete test sets without the need for a reference transcription. We also introduced a ``no-box'' WER estimation approach (e-WER2) which does not need to have access to the ASR system.
A potential limitation of this work is the restriction to only one language, so for future work, we shall continue our investigation to estimate WER across different languages and multilingual ASR systems. We also plan to use e-WER for lattice \textit{n}-best ranking 
for second pass rescoring.  

\smallskip
\noindent
\textbf{Acknowledgements:} This work was partially supported by EU H2020 project ``European Language Grid'' (grant agreement ID: 825627).

\bibliographystyle{IEEEtran}

\bibliography{ref}

\end{document}